\documentclass[showpacs,preprintnumbers,amsmath,amssymb]{revtex4}
\usepackage{amsmath}
\usepackage{graphicx}
\usepackage{subfig}
\usepackage{hyperref}
\usepackage{amssymb}
\usepackage[english]{babel}
\usepackage{epsfig}
\usepackage{wasysym}

\begin{document}

\title{ Born-Infeld and Charged Black Holes with non-linear source in $f(T)$ Gravity}

\author{ Ednaldo L. B. Junior$^{(a)}$\footnote{E-mail address:ednaldobarrosjr@gmail.com}, Manuel E. Rodrigues$^{(a,b)}$\footnote{E-mail address: esialg@gmail.com} and Mahouton J. S. Houndjo$^{(c,d)}$\footnote{E-mail address: sthoundjo@yahoo.fr} }
\affiliation{$^{a}$ \, Faculdade de F\'{\i}sica, PPGF, Universidade Federal do Par\'{a}, 66075-110, Bel\'{e}m, Par\'{a}, Brazil.\\
$^b$\ Faculdade de Ci\^{e}ncias Exatas e Tecnologia, Universidade Federal do Par\'{a}\\
Campus Universit\'{a}rio de Abaetetuba, CEP 68440-000, Abaetetuba, Par\'{a}, Brazil.\\
$^c$\ Institut de Math\'{e}matiques et de Sciences Physiques (IMSP),  
01 BP 613, Porto-Novo, B\'{e}nin.\\
$^d$ \, Facult\'{e} des Sciences et Techniques de Natitingou - Universit\'{e} de Parakou - B\'{e}nin.}


\begin{abstract}
We investigate $f(T)$ theory  coupled with a nonlinear source of electrodynamics, for a spherically symmetric and static  spacetime in $4D$. We re-obtain the Born-Infeld and Reissner-Nordstrom-AdS solutions. We generalize the no-go theorem for any content that obeys the relationship $\mathcal{T}^{\;\;0}_{0}=\mathcal{T}^{\;\;1}_{1}$ for the energy-momentum tensor and a given set of tetrads. Our results show new classes of solutions where the metrics are related through $b(r)=-Na(r)$. We do the introductory analysis showing that solutions are that of asymptotically flat black holes, with a singularity at the origin of the radial coordinate, covered by a single event horizon. We also reconstruct the action for this class of solutions and obtain the functional form $f(T)=f_0\left(-T\right)^{(N+3)/[2(N+1)]}$ and $\mathcal{L}_{NED}=\mathcal{L}_0\left(-F\right)^{(N+3)/[2(N+1)]}$. Using the Lagrangian density of Born-Infeld, we obtain a new class of charged black holes where the action reads  $f(T)=-16\beta_{BI}\left[1-\sqrt{1+(T/4\beta_{BI})}\right]$.
\end{abstract}


\maketitle



\section{Introduction}
\label{sec1}
Maxwell's electromagnetism is one of the most successful theories in the history of science. However, many other modifications of this theory have been made, and one of the best known arises from the Lagrangian density called Born-Infeld (BI) \cite{BI}. The initial idea, at $1937$, to build an electromagnetic theory for which the Lagrangian density is given as a non-linear function of the scalar formed by electric and magnetic fields, was to obtain solutions in which the value of the electric field, or magnetic, were finite at the origin where the electric charge is situated, unlike the case in which the Maxwell value diverges. For the BI solution, this property is satisfied, the value of the electric field is finite, considering only electrical charge. This theory still falls in Maxwell's one for the so-called weak field limit, where the BI parameter $ \beta_{BI}$ is taken for a limit tending to infinity.
\par 
Even in 1937, Hoffman and Infeld get a solution to this theory coupled to Einstein-Hilbert action \cite{hoffmann}. Only after several years this idea came to be explored again, with Peres \cite{peres}, so called non-linear electrodynamics (NED). A year later, Bardeen \cite{bardeen} suggests an exact solution of Einstein's equations in which there is an event horizon in the causal structure, but there is no singularity in all space, even at the origin of the radial coordinate. This solution is known as Bardeen ``regular'' black hole. Thus, Pellicer and Torrence get an exact solution of regular black hole where the asymptotic behavior is identical to the Reissner-Nordstrom's one \cite{pellicer}. With this, some other solutions also continued rising for NED \cite{plebanski1}.\par 
The NED continued to be addressed in many of its aspects \cite{NED}, particularly in cosmology \cite{novello}, where that it can be shown that the energy of a magnetic field can cause the acceleration of our universe at the present time.

\par 
On the other hand, General Relativity (GR) is also the theory that describes the gravitational interaction with the highest historically successful. Since the proof of star light bending passing near the sun region in 1919, the GR has proven very effective in the description of local phenomena and even cosmological. But for the current acceleration phase of our universe, shown by the most recent observations of type Ia supernovae \cite{Ia}, it can only describe the dynamics of the universe through an exotic fluid, commonly called dark energy, which has negative pressure. An alternative to try to improve this description, is the modification of Einstein's equations, or GR, for a further description made by an action different from that of Einstein-Hilbert. The main changes came from the Starobinski's suggestion where a term of second order (nonlinear) in the curvature scalar $R$ which appears in the action.  A famous theory that is now widely studied is the $f(R)$ Gravity \cite{fR}, where the action is given in terms of an analytic function of the curvature scalar, $S_{f(R)}=\int dx^4\sqrt{-g}[f(R)+2\kappa^2 \mathcal{L}_{Matter}]$. Some other changes exist, where the action is  taken  as a functional of nonlinear terms of a given scalar, $ f(R, \mathcal{T})$ Gravity \cite{fRT}, $ f(\mathcal{G})$ Gravity \cite{fG} and $f(R, \mathcal{G})$ \cite{fRG}.

\par 
Another possibility is to describe the gravitational interaction by considering the inertia of the motion. This possibility uses the torsion of the space-time, rather than the curvature, to account for the effects of inertia. Considering identically zero the Riemann tensor, thus assuming that the geometry is characterized only by the tetrads as dynamic fields, and the Weitzenbock connection as originating the torsion in the space-time, this would provide the same global and local phenomena of the GR, known as teleparallel theory (TT) \cite{TT,aldrovandi}. It is proven to be dynamically equivalent to GR, and the equations of motion of test particles are described by a force equation, which can be re-written as the famous equation of the geodesic in GR. In this theory the effects of inertia of the motion appear when the torsion tensor is non-zero, but the curvature is identically zero. Here the scalar built analogously to the curvature is the torsion scalar $T$. A possible direct generalization of this theory is known as $ f(T)$ Gravity  \cite{fT}, where the action is now given in terms of an analytic function of torsion scalar. Interestingly, this generalization came from a consideration of the action of a theory thus formulated, could be inspired by the NED, such as BI. The original work of Ferraro and Fiorini \cite{ferraro1} considers a generalization of TT in which the action is a functional of the torsion scalar with identical dependence of BI to the scalar $F$. This is one of the ideas that will be re-addressed here. In fact the soul of this work is to obtain black hole solutions using a Lagrangian density of the NED as material content. But this also will result in actions with identical terms to the NED for the torsion scalar.
\par 
Our main goal here is motivated by the great complexity of the equations of motion for  $f(T)$ theory,  when we include non-linear terms in analytical function $f(T)$. The equations of motion are written in such a way (non-linear differential coupled equations), for which it is almost impossible to get any specific solution. Thus, through the possibility of a NED, we can use a powerful method of solving these equations, assigning the new degrees of freedom for the electric field (or magnetic, by duality) to let the task more easy. We will make a detailed description of this method later.
\par
The paper is organized as follows. In section \ref{sec2} we perform the preliminary definitions and establish the equations of motion for the $f(T)$ Gravity for a NED as material content. In section \ref{sec3} we show that we can re-obtain BI solutions with cosmological constant, and the Reissner-Nordstrom-AdS as particular case where the BI parameter $\beta_{BI}$ is taken to infinity. In the section \ref{sec4} we have established a strong no-go theorem which greatly restricts the possibilities for obtaining solutions to the $f(T)$ Gravity. In section \ref{sec5} we obtain new solutions of black holes that do not violate the no-go theorem. Here they are divided into two great classes, in subsection \ref{subsec5.1} those who consider the relationship $b(r)=-Na(r)$, between the metric functions, and in subsection \ref{subsec5.2} those where $b(r)=Na(r)$. In subsection \ref{subsec5.3} we obtain a Born-Infeld-type solution for $b(r)\neq -a(r)$. We ended up with our final considerations in section \ref{sec6}.

\section{Non-linear electrodynamics source in $f(T)$ Gravity}
\label{sec2}

In a differentiable manifold ``$\mathcal{M}$" one can define a tangent space (or co-tangent) to a point $P$ of the same, as that which contains all the tangent vectors  to the manifold at this point, such that this set vectors obey the properties of a vector space. We denote the tangent space by $ \mathbb{T}_p(\mathcal{M})$. Thus we can generalize this definition to a vector field where the tangent vectors can be defined at any point $ P \in \mathcal{M}$, then we represent by $ \mathbb{T}(\mathcal{M})$. The dual space to the tangent is called the co-tangent space $\mathbb{T}^*(\mathcal{M})$. The latter contains all co-vectors in $\mathcal{M}$. Every element of $\mathbb{T}(\mathcal{M})$ can be written in the form $ V = V^{\mu} \partial_{\mu}$ ($\partial_{\mu}=\partial/\partial x^{\mu}$), and every element of $ \mathbb{T}^*(\mathcal{M})$ can be written as $\omega = \omega_{\mu}dx^{\mu}$, where $ \{\partial_{\mu}\}$ and $ \{dx^{\mu}\}$ are local basis linearly independent from the  tangent and cotangent spaces respectively.
\par 
The local basis  are related to the general basis by  $e_{a}=e^{\;\;\mu}_{a}\partial_{\mu}$ and $e^{a}=e^{a}_{\;\;\mu}dx^{\mu}$, where the matrices obey the relations $e^{a}_{\;\;\mu}e_{a}^{\;\;\nu}=\delta^{\nu}_{\mu}$ and $e^{a}_{\;\;\mu}e_{b}^{\;\;\mu}=\delta^{a}_{b}$. The general basis of the tangent space satisfy the following commutation relation
\begin{eqnarray}
[e_{a},e_{b}]=f^{c}_{\;\;bc}e_{c}\,,\label{cr1}
\end{eqnarray}
where $f^{c}_{\;\;bc}e_{c}$ are the coefficients of structure or anholonomy. Using the first equation of Cartan structure  ($de^{c}=-(1/2)f^{c}_{\;\;ab}e^{a}\wedge e^{b}$)  and the relationship between the local basis, one gets  $f^{c}_{\;\;ab}=e_{a}^{\;\;\mu}e_{b}^{\nu}(\partial_{\nu}e^{c}_{\;\;\mu}-\partial_{\mu}e^{c}_{\;\;\nu})$. We can then define an inertial reference frame when the structure coefficients are all identically zero, i.e,  $f^{c}_{\;\;ab}\equiv 0$. In general, for any one frame, we do not necessarily have this condition.
\par 
With this structure, the line element in  $\mathcal{M}$, it can be represented local and general basis as 
\begin{eqnarray}
dS^2=g_{\mu\nu}dx^{\mu}dx^{\nu}=\eta_{ab} e^{a}e^{b}\label{ele}\;,\\
e^{a}=e^{a}_{\;\;\mu}dx^{\mu}\;,\;dx^{\mu}=e_{a}^{\;\;\mu} e^{a}\label{the}\;,
\end{eqnarray}
where $g_{\mu\nu}$ is the metric of the space-time, $\eta_{ab}$ the Minkowski metric. The root of the determinant of the metric is given by $\sqrt{-g}=det[e^{a}_{\;\;\mu}]=e$. A possible special structure is the one for which  we demand that the Riemann tensor vanishes identically, which is obtained with the following Weitzenbok connection
\begin{eqnarray}
\Gamma^{\sigma}_{\;\;\mu\nu}=e_{a}^{\;\;\sigma}\partial_{\nu}e^{a}_{\;\;\mu}=-e^{a}_{\;\;\mu}\partial_{\nu}e_{a}^{\;\;\sigma}\label{co}\; .
\end{eqnarray}
\par
Through the connection, we can define the components of the torsion and  contorsion tensors  as
\begin{eqnarray}
T^{\sigma}_{\;\;\mu\nu}&=&\Gamma^{\sigma}_{\;\;\nu\mu}-\Gamma^{\sigma}_{\;\;\mu\nu}=e_{a}^{\;\;\sigma}\left(\partial_{\mu} e^{a}_{\;\;\nu}-\partial_{\nu} e^{a}_{\;\;\mu}\right)\label{tor}\;,\\
K^{\mu\nu}_{\;\;\;\;\alpha}&=&-\frac{1}{2}\left(T^{\mu\nu}_{\;\;\;\;\alpha}-T^{\nu\mu}_{\;\;\;\;\alpha}-T_{\alpha}^{\;\;\mu\nu}\right)\label{cont}\; .
\end{eqnarray}
\par
For facilitating the description of the Lagrangian density and the equations of motion, we can define another tensor from the components of the torsion and contorsion tensors, as
\begin{eqnarray}
S_{\alpha}^{\;\;\mu\nu}=\frac{1}{2}\left( K_{\;\;\;\;\alpha}^{\mu\nu}+\delta^{\mu}_{\alpha}T^{\beta\nu}_{\;\;\;\;\beta}-\delta^{\nu}_{\alpha}T^{\beta\mu}_{\;\;\;\;\beta}\right)\label{s}\;.
\end{eqnarray}
We can then define the scale of this theory, the torsion scalar, as follows
\begin{equation}
T=T^{\alpha}_{\;\;\mu\nu}S_{\alpha}^{\;\;\mu\nu}\label{t1}\,.
\end{equation}
\par 
We will establish the equations of motion for the case of a source of non-linear electrodynamics. First we take the following Lagrangian density
\begin{eqnarray}
\mathcal{L}=e\left[f(T)+2\kappa^2\mathcal{L}_{NED}(F)\right],\label{lagrangean}
\end{eqnarray} 
where $\mathcal{L}_{NED}(F)$ is the contribution of non-linear electrodynamics (NED), with $F=(1/4)F_{\mu\nu}F^{\mu\nu}$ being  $F_{\mu\nu}$ the components the Maxwell's tensor, and  $\kappa^2=8\pi G/c^4$, where $G$ is the Newtonian constant and  ``$c$'' the speed of light. To establish the equations of motion by Euler-Lagrange ones, we have to take the derivations with respect to the tetrads

\begin{eqnarray}
&&\frac{\partial\mathcal{L}}{\partial e^{a}_{\;\;\mu}}=f(T)ee_{a}^{\;\;\mu}+ef_{T}(T)4e^{a}_{\;\;\alpha}T^{\sigma}_{\;\;\nu\alpha}S_{\sigma}^{\;\;\mu\nu}+2\kappa^2\frac{\partial\mathcal{L}_{NED}}{\partial e^{a}_{\;\;\mu}}\,,\\
&&\partial_{\alpha}\left[\frac{\partial\mathcal{L}}{\partial (\partial_{\alpha}e^{a}_{\;\;\mu})}\right]=-4f_{T}(T)\partial_{\alpha}\left(ee_{a}^{\;\;\sigma}S_{\sigma}^{\;\;\mu\nu}\right)-4ee_{a}^{\;\;\sigma}S_{\sigma}^{\;\;\mu\gamma}\partial_{\gamma}T f_{TT}(T)+2\kappa^2\partial_{\alpha}\left[\frac{\partial\mathcal{L}_{NED}}{\partial (\partial_{\alpha}e^{a}_{\;\;\mu})}\right]\,,
\end{eqnarray}
with $f_{T}(T)=df(T)/dT$ and  $f_{TT}(T)=d^2f(T)/dT^2$. With the  above expressions and  Euler-Lagrange's equations
\begin{eqnarray}
\frac{\partial\mathcal{L}}{\partial e^{a}_{\;\;\mu}}-\partial_{\alpha}\left[\frac{\partial\mathcal{L}}{\partial (\partial_{\alpha}e^{a}_{\;\;\mu})}\right]=0\label{ELeq}\,,
\end{eqnarray} 
and multiplying by the  factor $e^{-1}e^{a}_{\;\;\beta}/4$, one gets the following equations of motion
\begin{eqnarray}
S_{\beta}^{\;\;\mu\alpha}\partial_{\alpha}T f_{TT}(T)+\left[e^{-1}e^{a}_{\;\;\beta}\partial_{\alpha}\left(ee_{a}^{\;\;\sigma}S_{\sigma}^{\;\;\mu\alpha}\right)+T^{\sigma}_{\;\;\nu\beta}S_{\sigma}^{\;\;\mu\nu}\right]f_{T}(T)+\frac{1}{4}\delta^{\mu}_{\beta}f(T)=\frac{\kappa^2}{2}\mathcal{T}^{\;\;\mu}_{\beta}\label{meq}
\end{eqnarray} 
where $\mathcal{T}_{\beta}^{\;\;\mu}$ is the energy-momentum tensor of of the source of non-linear electrodynamics 
\begin{eqnarray}
\mathcal{T}_{\beta}^{\;\;\mu}=-\frac{2}{\kappa^2}\left[\delta^{\mu}_{\beta}\mathcal{L}_{NED}(F)-\frac{\partial \mathcal{L}_{NED}(F)}{\partial F}F_{\beta\sigma}F^{\mu\sigma}\right]\label{emtensor}\,.
\end{eqnarray} 
Because of the relationship in (\ref{ele}), we can choose several frames that fall into a spherically  symmetric and static  metric, in spherical coordinates. Also the various matrices tetrads are connected by a Lorentz transformation. Taking a diagonal matrix $[\bar{e}^{a}_{\;\;\mu}]=diag[e^{a(r)},e^{b(r)},r,r\sin\theta]$, to be the dynamic field theory, but it falls on a inconsistency in the equations of motion (\ref{meq}), appearing as a spurious component $\theta-r$ ($2-1$). This is precisely due to the proper choice of reference for this symmetry, as seen in \cite{tamanini}. Therefore, we undertake a good choice given in \cite{rodrigues3}, which is simply a rotation by a specific Lorentz transformation of the diagonal case, where all the equations appear consistent, and  described by
\begin{eqnarray}\label{ndtetrad}
\{e^{a}_{\;\;\mu}\}=\left[\begin{array}{cccc}
e^{a/2}&0&0&0\\
0&e^{b/2}\sin\theta\cos\phi & r\cos\theta\cos\phi &-r\sin\theta\sin\phi\\
0&e^{b/2}\sin\theta\sin\phi &
r\cos\theta\sin\phi &r\sin\theta
\cos\phi  \\
0&e^{b/2}\cos\theta &-r\sin\theta  &0
\end{array}\right]\;,
\end{eqnarray}  
where, through the relationship given in (\ref{ele}), we can reconstruct the metric as being 
\hspace{0,2cm}
\begin{equation}
dS^2=e^{a(r)}dt^2-e^{b(r)}dr^2-r^2\left[d\theta^{2}+\sin^{2}\left(\theta\right)d\phi^{2}\right]\label{ltb}\;,
\end{equation}
in which the metric functions $\{a(r),b(r)\}$ are assumed to be functions of radial coordinate $r$ and are not time dependent. 
\par
With this,  one can now calculate all the geometrical objects established in the theory. The non-zero components of the torsion tensor (\ref{tor}) are
\begin{eqnarray}
T^{0}_{\;\;10}=\frac{a'}{2}\,,\;\;\;T^{2}_{\;\;21}=T^{3}_{\;\;31}=\frac{e^{b/2}-1}{r}\,\,,\label{tt}
\end{eqnarray}
while the non-null components of the contorsion tensor read
\begin{eqnarray}
K_{\;\;\;\;0}^{10}=\frac{a'e^{-b}}{2}\,,\;\;\;K_{\;\;\;\;1}^{22}=K_{\;\;\;\;1}^{33}=\frac{e^{-b}(e^{b/2}-1)}{r}\,\,.\label{ctt}
\end{eqnarray}
We also calculate the non-null components of the tensor  $S_{\alpha}^{\;\;\mu\nu}$, as
\begin{eqnarray}
S_{0}^{\;\;01}=\frac{e^{-b}(e^{b/2}-1)}{r}\,,\;\;S_{2}^{\;\;12}=S_{3}^{\;\;13}=\frac{e^{-b}\left(a'r-2e^{b/2}+2\right)}{4r}\,.\label{st}
\end{eqnarray}
\par
From the definition of the torsion scalar (\ref{t1}), one gets
\begin{equation}
T=\frac{2}{r}\left[-\left(a'+\frac{2}{r}\right)e^{-b/2}+\left(a'+\frac{1}{r}\right)e^{-b}+\frac{1}{r}\right]  \label{te}\,.
\end{equation}
We note here that in general, the torsion scalar is an arbitrary function of the radial coordinate $r$.
\par
Let us consider here that there is only electric charge, so we can set the electric four-potential as a 1-form $A=A_{\mu}dx^{\mu}$, where the components are given by $\{A_{\mu}\}=\{A_{0}(r),0,0,0\}$ and  $A_{0}$  is the electric scalar potential. We define the Maxwell tensor as a 2-form $F=(1/2)F_{\mu\nu}dx^{\mu}\wedge dx^{\nu}$, where the components are anti-symmetric and given by $F_{\mu\nu}=\nabla_{\mu}A_{\nu}-\nabla_{\nu}A_{\mu}$. From our restriction of the electric scalar potential, the Maxwell tensor has only one independent component $F_{10}(r)$. This can still be seen by the spherical symmetry of space-time. As the metric has spherical symmetry, the equation that defines the vectors of Killing vectors $\mathcal{L}_{K^{\sigma}}g_{\mu\nu}(r,\theta)=0$,  leading to vectors $\{K^{\sigma}\}$ being the generators of the $O(3)$ group, yielding  $\mathcal{L}_{K^{\sigma}} F^{\mu\nu}(t,r,\theta)=0$. This restricts us to only two non-null independent components of the Maxwell field
$F^{10}(t,r)$ and  $F^{23}(t,r,\theta)$. Considering only electric source, one gets from the equations of motion with respect to  $A^{\mu}$, and that the metric is also static, where there exists only the component $F^{10}(r)$ \cite{wainwright}.  Making use of  Euler-Lagrange equations for the field $ A_{\mu}$, for the Lagrangian density (\ref{lagrangean}), we have the modified Maxwell equation
\begin{equation}
\nabla _{\mu}\left[F^{\mu\nu}\frac{\partial\mathcal{L}_{NED}}{\partial F}\right]=\partial_{\mu}\left[e^{-1}F^{\mu\nu}\frac{\partial\mathcal{L}_{NED}}{\partial F}\right]=0\label{mMaxeq}\,,
\end{equation}
whose solution, using  $\nu=0$, is given by
\begin{eqnarray}
F^{10}(r)=\frac{q}{r^2}e^{-(a+b)/2}\left(\frac{\partial\mathcal{L}_{NED}}{\partial F}\right)^{-1}\,,\label{F10}
\end{eqnarray}
where $q$ is the electric charge.
\par
Through the components of the torsion scalar  (\ref{tt}), of (\ref{st}), (\ref{te}) and the energy-momentum  tensor (\ref{emtensor}), the equations of motion (\ref{meq}) become
\begin{eqnarray}
&&2\frac{e^{-b}}{r}\left(e^{b/2}-1\right)T'f_{TT}+\frac{e^{-b}}{r^2}\left[b'r+\left(e^{b/2}-1\right)\left(a'r+2\right)\right]f_T+\frac{f}{2}=-2\mathcal{L}_{NED}-2\frac{q^2}{r^4}\left(\frac{\partial\mathcal{L}_{NED}}{\partial F}\right)^{-1}\label{eq1}\,,\\
&&\frac{e^{-b}}{r^2}\left[\left(e^{b/2}-2\right)a'r
+2\left(e^{b/2}-1\right)\right]f_T+\frac{f}{2}=-2\mathcal{L}_{NED}-2\frac{q^2}{r^4}\left(\frac{\partial\mathcal{L}_{NED}}{\partial F}\right)^{-1}\label{eq2}\,,\\
&&\frac{e^{-b}}{2r}\left[a'r+2\left(1-e^{b/2}\right)\right]T'f_{TT}+
\frac{e^{-b}}{4r^2}\Big[\left(a'b'-2a''-a'^2\right)r^2+\nonumber\\
&&+\left(2b'+4a'e^{b/2}-6a'\right)r-
4e^{b}+8e^{b/2}-4\Big]f_{T}+\frac{f}{2}=-2\mathcal{L}_{NED}\label{eq3}\,.
\end{eqnarray}
\par 
In the next section we will show how to re-obtain some of the well known  solutions of the GR, arising from the matter content proposed here. 

\section{Recovering Born-Infeld and some known solutions}\label{sec3}

Here we will show that the theory described above leads to all the known solutions of the GR coupled to nonlinear electrodynamics. Let us perform here some best known cases, starting with the Lagrangian density of Born-Infeld (BI). The Lagrangian density is given by BI
\begin{eqnarray}
\mathcal{L}_{NED}=\mathcal{L}_{BI}=4\beta_{BI}^2\left[1-\sqrt{1+\frac{F}{2\beta_{BI}^2}}\right]\,,\label{lBI}
\end{eqnarray} 
with $F=(1/4)F_{\mu\nu}F^{\mu\nu}$ and  $\beta_{BI}$ being the BI parameter. This Lagrangian density falls into the case of Maxwell electromagnetic theory, for the limit where the BI parameter tends to infinity, then
\begin{eqnarray}
\lim_{\beta_{BI}\rightarrow \infty}\mathcal{L}_{BI}\sim -F+O\left[F^2\right]\sim \mathcal{L}_{Maxwell}\,.\label{lMaxwell}
\end{eqnarray} 
Let us look at the particular cas of the TT, where $f(T)=T-2\Lambda,f_T(T)=1$ and  $f_{TT}(T)=0$. For the BI Lagrangian density  (\ref{lBI}) the equations of motion (\ref{eq1})-(\ref{eq3}) lead to 
\begin{eqnarray}
&&e^{-b}\left(1-rb'\right)=1-r^2\Lambda+2\left[r^2\mathcal{L}_{BI}+\frac{q^2}{r^2}\left(\frac{\partial \mathcal{L}_{BI}}{\partial F}\right)^{-1}\right]\label{eq1BI}\,,\\
&& e^{-b}\left(1+ra'\right)=1-r^2\Lambda+2\left[r^2\mathcal{L}_{BI}+\frac{q^2}{r^2}\left(\frac{\partial \mathcal{L}_{BI}}{\partial F}\right)^{-1}\right]\label{eq2BI} \,,\\
&& e^{-b}\left[ 2a''+a'^{2}-\frac{b'}{r}(2+ra')+2\frac{a'}{r}\right]=-4\Lambda +8\mathcal{L}_{BI}\,.
\end{eqnarray}
The general property for spherical symmetry appears by subtracting the equation  (\ref{eq2BI}) from (\ref{eq1BI}), yielding 
\begin{eqnarray}
re^{-b}\left(a'+b'\right)=0\,.\label{ssimmetry}
\end{eqnarray}
Without loss of generality, we use $b(r)=-a(r)$ in (\ref{eq2BI}), recalling (\ref{F10}), we have 
\begin{eqnarray}
\left(re^{a}\right)^{\prime}=1-\Lambda r^2+4\beta_{BI}\left(2\beta_{BI}r^2-\sqrt{q^2+4\beta_{BI}^2r^4}\right)
\end{eqnarray}
which, after integration, gives the following solution 
\begin{eqnarray}
&&e^{a(r)}=e^{-b(r)}=1-\frac{2M}{r}-\frac{\Lambda}{3}r^2+4\beta_{BI}\left(2\frac{\beta_{BI}}{3}r^2-\frac{1}{r}\int^{r}_{\infty}\sqrt{q^2+4\beta_{BI}^2r^4}dr\right)\label{sBI1}\,,\\
&&F^{10}(r)=\frac{2\beta_{BI}q}{\sqrt{q^2+4\beta_{BI}^2r^4}}\label{sBI2}\,.
\end{eqnarray}
This is the solution analogous to the one of  Einstein-Born-Infeld with cosmological constant obtained in  \cite{fernando1}\footnote{ Note that the convention of  Fernando et al. is given by  $F=F_{\mu\nu}F^{\mu\nu}$, different from ours where $F=(1/4)F_{\mu\nu}F^{\mu\nu}$. This shows that $\beta_{BI}$ is the half of theirs.}.   
\par
Taking now the particular case where  $\beta_{BI}$ goes to infinity, the Lagrangian density  $\mathcal{L}_{NED}=\mathcal{L}_{Maxwell}=-F$ and  $\partial\mathcal{L}_{NED}/\partial F=-1$. Using (\ref{F10}) in the equations of motion (\ref{eq1})-(\ref{eq3}), one gets
\begin{eqnarray}
&&e^{-b}\left(1-rb'\right)=1-r^2\Lambda-\frac{q^2}{r^2}\label{eq1M}\,,\\
&& e^{-b}\left(1+ra'\right)=1-r^2\Lambda-\frac{q^2}{r^2}\label{eq2M} \,,\\
&& e^{-b}\left[ 2a''+a'^{2}-\frac{b'}{r}(2+ra')+2\frac{a'}{r}\right]=-4\Lambda +\frac{q^2}{r^4}\,.
\end{eqnarray}
Once again we subtract (\ref{eq2M}) from (\ref{eq1M}), yielding $b(r)=-a(r)$. Using this result in  (\ref{eq2M}) and integrating, one gets the  Reissner-Nordstrom-AdS solution  (RN-dS for $\Lambda>0$)
\begin{eqnarray}
&&e^{a(r)}=e^{-b(r)}=1-\frac{2M}{r}+\frac{q^2}{r^2}-\frac{\Lambda}{3}r^2\,,\,F^{10}(r)=-\frac{q}{r^2}\label{sM}\,.
\end{eqnarray}
This solution has been re-obtained in  \cite{rodrigues1}. All the particular cases, with $\Lambda=0$ RN, with $q=0$ S-dS and  S-AdS also are re-obtained here.

\section{No-Go Theorem for Born-Infeld-type and general charged solutions in f(T) Gravity}\label{sec4}


In some cases models of charged black holes appear certain restrictions that prevent obtaining new solutions to certain conditions. These restrictions are commonly called ``no-go theorem''. The restriction in the radial and azimuthal components of the electric field can not be simultaneously zero appears in models with $(1+2)D$, in the GR \cite{cataldo2}, and with curvature and torsion in \cite{blagojevic}. In the $f(T)$ Gravity, in $(1+2)D$ we have in \cite{gonzalez}, and for $d$-dimensions in \cite{capozziello1}.
\par 
We will show now that the solutions of type $b(r)=-a(r)$ can exist only in the particular case of the TT. To do this, we will subtract the equation (\ref{eq2}) from (\ref{eq1}), yielding after rearrangement 
\begin{eqnarray}
\frac{d}{dr}\left[\ln f_T(T)\right]+\left[\frac{a'+b'}{2\left(e^{b/2}-1\right)}\right]=0\label{no-go}\,.
\end{eqnarray}
In this case, we can enunciate the theorem as follows:
\par 
{\bf Theorem}: The solutions of a Weitzenbock spacetime with spherical and static symmetry whose the sets of tetrads are given  (\ref{ndtetrad}), with the restriction $b(r)=-a(r)$ and $\mathcal{T}^{\;\;0}_{0}=\mathcal{T}^{\;\;1}_{1}$ in (\ref{meq}), exist if and only if the theory is Teleparallel, where $f(T)$ is a linear function of the torsion scalar.
\par 
{\bf Proof}. According to the equation (\ref{no-go}), we can only get the following situations to restriction  $b(r)=-a(r)$:
\begin{enumerate}
\item When $b(r)=-a(r)$, from (\ref{no-go}) one gets $\left(\ln f_T(T)\right)^{\prime}=0$, yield  $f_T(T)=a_1$, with  $a_1$ being a real constant. This lead to the solution $f(T)=a_1T+a_0$ linear in $T$, with $a_0$ another real constant;\par  
\item When $f_T(T_0)=0$, with the torsion being a real constant $T=T_0$. In this case, the equation (\ref{no-go}) leads directly to the restriction $b(r)=-a(r)$ and $f(T_0)=f_0$ with $f_0$ real constant;\par    
\item When $f_T(T_0)\neq 0$, with $T_0$ being a real constant, the equation (\ref{no-go}) restricts to $b(r)=-a(r)$ and $f(T_0)=f_0$ as before.
\end{enumerate}
Note that this theorem is valid not only for general solutions with non-linear source of electrodynamics, but for whatever the energy-momentum tensor satisfying the required conditions, i.e, for every theory whose the set of tetrads are given by (\ref{ndtetrad}), with the restriction $ b(r)=-a(r)$ and $\mathcal{T}^{\;\;0}_{0}=\mathcal{T}^{\;\;1}_{1}$ in  in (\ref{meq}).
\par
So there are only solutions with spherical and static symmetry, with the usual tetrads (\ref{ndtetrad}), for models with matter described through an anisotropic energy-momentum tensor, where $\mathcal{T}^{\;\;0}_{0}\neq\mathcal{T}^{\;\;1}_{1}$. We have a case similar to the famous symmetry $[(d^2/dtdr)A(r,t)]B(r,t)=[(d/dr)A(r,t)][(d/dt)B(r,t)]$ of Lema\^{\i}tre-Tolman-Bondi models. If the spherically symmetric  metric is described by a set of diagonal tetrads with an anisotropic energy-momentum tensor, the symmetry implies that the function $f(T)$ is linear in $T$. But when the set of tetrads is non-diagonal, this symmetry can be generalized or broken down when the analytical function  $f(T)$ has  nonlinear corrections in $T$  (see the third expression of equation (42) \cite{rodrigues2}). So in that sense, there is also a no-go theorem to maintain this symmetry in the models of Lema\^{\i}tre-Tolman-Bondi.   

\section{New charged solutions with non-linear source}\label{sec5}


\subsection{Solutions with $b(r)=-Na(r)$}\label{subsec5.1}
As the equations of motion (\ref{eq1})-(\ref{eq3}) are a highly complicated equations, we will establish here a powerful algorithm to solve it. In the case of nonlinear electrodynamics, we have a crucial alternative to solve these equations, which is the freedom of the choice Lagrangian density model $\mathcal{L}_{NED}$. We consider here all functions as only dependent coordinate $r$. So, let's start our algorithm by integrating equation (\ref{no-go}) in the previous section, which leads to 
\begin{eqnarray}
f_T(r)=\exp\left\{-\int \left[\frac{a'+b'}{2\left(e^{b/2}-1\right)}\right]dr\right\}\label{fT}.
\end{eqnarray}   
According to the result of the previous section, the no-go theorem, we do not demand the particular case $b=-a$ directly. Now we take the cases where  
\begin{eqnarray}
b(r)=-Na(r)\;,\; a(r)=\log\left[1-\frac{2M(r)}{r}\right]\,,\label{b=-Na} 
\end{eqnarray}
with $N\in\mathbb{N}$ and  $M(r)$ an arbitrary function of the radial coordinate $r$. This does not fix the solution, but only establish the relation between the metric functions  $a(r)$ and  $b(r)$. Note that the particular case  $b=-a$ is include here. We solve the equation  (\ref{eq1}) for determine  $\mathcal{L}_{NED}$
\begin{eqnarray}
&&\mathcal{L}_{NED}(r)=\frac{1}{2}\Bigg\{-\frac{1}{2}f(r)+\frac{2M(r)}{r^4\mathcal{L}_F(r-2M(r))}\Bigg[2q^2+r^2\mathcal{L}_F\left(1-\frac{2M(r)}{r}\right)^{N/2}\nonumber\\
&&\exp\left(\int\frac{(N-1)\left(1-\frac{2M(r)}{r}\right)^{N/2}(rM'(r)-M(r))}{r(r-2M(r))
\left(-1+\left(1-\frac{2M(r)}{r}\right)^{N/2}\right)}dr\right)\Bigg]-\frac{2}{r^3\mathcal{L}_F(r-2M(r))}\nonumber\\
&&\Bigg[q^2+r^2\mathcal{L}_F\left(1-\frac{2M(r)}{r}\right)^{N/2}\exp\left(\int\frac{(N-1)\left(1-\frac{2M(r)}{r}\right)^{N/2}(rM'(r)-M(r))}{r(r-2M(r))
\left(-1+\left(1-\frac{2M(r)}{r}\right)^{N/2}\right)}dr\right)\nonumber\\
&&\Bigg(1-\left(1-\frac{2M(r)}{r}\right)^{N/2}+\left(-1+2\left(1-\frac{2M(r)}{r}\right)^{N/2}\right)M'(r)\Bigg)\Bigg]\Bigg\}\label{L}\,,
\end{eqnarray}
where $\mathcal{L}_F=\partial \mathcal{L}_{NED}/\partial F$ and  $M'(r)=dM(r)/dr$. Subtracting the equation  (\ref{eq3}) from  (\ref{eq2}), we get $\mathcal{L}_F$
\begin{eqnarray}
&&\mathcal{L}_F=\Bigg\{2q^2(r-2M(r))^2\left[-1+\left(1-\frac{2M(r)}{r}\right)^{N/2}\right]\nonumber\\
&&\exp\left(-\int\frac{(N-1)\left(1-\frac{2M(r)}{r}\right)^{N/2}(rM'(r)-M(r))}{r(r-2M(r))
\left(-1+\left(1-\frac{2M(r)}{r}\right)^{N/2}\right)}dr\right)\Bigg\}\Bigg/ \Bigg\{r^2\Bigg[M^2(r)\Bigg(4-4\left(1-\frac{2M(r)}{r}\right)^{N/2}\nonumber\\
&&+3\left(1-\frac{2M(r)}{r}\right)^{N}-2\left(1-\frac{2M(r)}{r}\right)^{3N/2}+N\left(1-\frac{2M(r)}{r}\right)^{N}\left(-3+2\left(1-\frac{2M(r)}{r}\right)^{N/2}\right)\Bigg)\Bigg]\nonumber\\
&&r^2\Bigg[2M'(r)(N-2)\left(1-\frac{2M(r)}{r}\right)^{N}\left(-1+\left(1-\frac{2M(r)}{r}\right)^{N/2}\right)-(N-1)M'^2(r)\left(1-\frac{2M(r)}{r}\right)^{N}\nonumber\\
&&\left(-1+2\left(1-\frac{2M(r)}{r}\right)^{N/2}\right)+\left(-1+\left(1-\frac{2M(r)}{r}\right)^{N/2}\right)\Bigg(-1+\left(1-\frac{2M(r)}{r}\right)^{N}\nonumber\\
&&+r\left(1-\frac{2M(r)}{r}\right)^{N}M''(r)\Bigg)\Bigg]-2rM(r)\Bigg[-\left(1-\frac{2M(r)}{r}\right)^{N}\left(N-3+2\left(1-\frac{2M(r)}{r}\right)^{N/2}\right)M'(r)\nonumber\\
&&+\left(-1+\left(1-\frac{2M(r)}{r}\right)^{N/2}\right)\left(-2+N\left(1-\frac{2M(r)}{r}\right)^{N}+r\left(1-\frac{2M(r)}{r}\right)^{N}M''(r)\right)\Bigg]\Bigg\}\,.\label{LF}
\end{eqnarray}
Thus, with the expressions of $ \mathcal{L}_{NED}$ and $ \mathcal{L}_F$  in (\ref{L}) and (\ref{LF}), the function (\ref{fT}) and the relation (\ref{b=-Na}), all the equations of motion (\ref{eq1})-(\ref{eq3}) are satisfied. The point is that these expressions are not necessarily related to $ \mathcal{L}_F$ which is the derivative of $ \mathcal{L}_{NED}$ with respect to  $F$. These expressions were obtained to satisfy the equations of motion, leaving them free and unrelated to each other. Just below, we will formulate a method which requires the correct relationship between these expressions for the solution obtained being consistent. This forces a specific solution for each particular case for the function $M (r)$.
\par   
Now let's specify some particular cases for the values of $ N $.
\subsubsection{b(r)=-2a(r)}
Taking $N=2$, one has the particular case $b(r)=-2a(r)$, which, after integration of the equation (\ref{fT}) leads to 
\begin{eqnarray}
f(r)=8\int\exp\left[\int\frac{1}{2}\left(\frac{1}{r}-\frac{M'(r)}{M(r)}\right)dr\right]\frac{1}{r^4}\left[rM'^2(r)+M(r)(rM''(r)-3M'(r))\right]dr\label{f}\,.
\end{eqnarray}  
Using (\ref{LF}) in (\ref{F10}), one gets 
\begin{eqnarray}
F^{10}(r)&=&\frac{1}{4qr^2M(r)}\sqrt{1-\frac{2M(r)}{r}}\Bigg\{4M^3(r)+r^3M'^2(r)+rM^2(r)\left(8M'(r)-4rM''(r)-7\right)\nonumber\\
&&+2r^2M(r)\left(rM''(r)-2M'^2(r)-M'(r)\right)\Bigg\}\exp\left[\frac{1}{2}\int\left(\frac{1}{r}-\frac{M'(r)}{M(r)}\right)dr\right]\label{F10-0}
\end{eqnarray}
Note here that there exists only an unique free function, the $M(r)$, in the expressions (\ref{f}) and  (\ref{F10-0}).
\par 
It is known that the derivative of the matter Lagrangian density with respect to  $F$, i.e, $\mathcal{L}_F$ in (\ref{LF}) for $N=2$, has to satisfy  
\begin{eqnarray}
\mathcal{L}_F=\frac{\partial \mathcal{L}_{NED}}{\partial r}\left(\frac{dF}{dr}\right)^{-1}\label{imposition1}\,.
\end{eqnarray}
From this relation, and using (\ref{L}), we have the following relationship equation
\begin{eqnarray}
r^2M'^2(r)-M^2(r)=0\,,\label{si1}
\end{eqnarray}
which leads to $M(r)=c_0r$, which causes an indetermination in equation (\ref{imposition1}), and then we discard this solution. Another possible solution of (\ref{si1}) is
\begin{eqnarray}
M(r)=\frac{r_0}{r}\;,\;r_0\in\Re_{+}\label{M1}\,.
\end{eqnarray} 
Remembering that (\ref{b=-Na}) and (\ref{F10-0}), then we have the solution following 
\begin{eqnarray}
&&dS^2=\left(1-\frac{2r_0}{r^2}\right)dt^2-\left(1-\frac{2r_0}{r^2}\right)^{-2}dr^2-r^2\left(d\theta ^2+\sin^2\theta d\phi^2\right)\label{s1}\,,\\
&&F^{10}(r)=-\frac{4r_0^2}{qr^3}\sqrt{1-\frac{2r_0}{r^2}}\label{F10-1}\,.
\end{eqnarray}

We can reconstruct the Lagrangian density $\mathcal{L}_{NED}$ and the algebraic function $f(T)$. From (\ref{s1}) and (\ref{te}), we get  
\begin{eqnarray}
T(r)=-\frac{8r_0^2}{r^6}\label{te1}\,.
\end{eqnarray}
Inverting this expression for obtaining $r(T)$ and using (\ref{s1}) into (\ref{f}), we get 
\begin{eqnarray}
f(T)=-\frac{6}{5}\sqrt{2}r_0^{1/3}\left(-T\right)^{5/6}\,.\label{f1}
\end{eqnarray}
On the same way, using (\ref{s1}) and (\ref{F10-1}) one gets 
\begin{eqnarray}
F(r)=-\frac{8r_0^4}{q^2r^6}\label{F1}\,.
\end{eqnarray}
By inverting this expression for obtaining  $r(F)$ and using  $f(r)$ and  (\ref{LF}) into (\ref{L}), one gets
\begin{eqnarray}
\mathcal{L}_{NED}=\frac{3q^{5/3}}{5\sqrt{2}r_0^4}\left(-F\right)^{5/6}\label{L1}\,.
\end{eqnarray}
This is the first case and we can note that the exponent of the torsion scalar  $T$, into $f(T)$, is identical to the scalar  $F$, in $\mathcal{L}_{NED}$.
\par 
The curvature scalar  $R=g^{\mu\nu}g^{\alpha\beta}R_{\alpha\mu\beta\nu}$, where  $R_{\alpha\mu\beta\nu}$ are the components of the Riemann tensor, can be obtained from the relation  
\begin{eqnarray}
R=-T-2\nabla^{\mu}T^{\nu}_{\;\;\mu\nu}\,,
\end{eqnarray}
and the components  (\ref{tt}), resulting for (\ref{s1})
\begin{eqnarray}
R(r)=\frac{4r_0}{r^6}(r^2-2r_0)\label{R1}\,.
\end{eqnarray}

We can then make an introductory analysis of the solution in (\ref{s1}) and (\ref{F10-1}). We see that from the definition of  the event horizon for a spherical symmetry of a static space-time \cite{plebanski}, we require $g^{11}(r_H)=0$, resulting to our case $r_H= \sqrt{2r_0}$. The metric is asymptotically Minkowskian. $g_{00}>0$ for $r^2>2r_0$, $g_{00}=0$ into $r=r_H=\sqrt{2r_0}$ and  $g_{00}<0$ for  $r^2<2r_0$. $g_{11}(r)$ is always positive. The component $F^{10}(r)$ in (\ref{F10-1}) behaves as  $F^{10}\sim r^{-3}$ in the limit  $r\rightarrow +\infty$.  This behavior was only seen atin  $5D$ in the  GR in \cite{ortaggio}. Then the Maxwell field vanishes at spatial infinity, keeping the characteristic of asymptotically flat. This component vanishes on the horizon $ r= r_H$. The curvature scalar (\ref{R1}) vanishes in the infinite space and diverges at the origin $ r = 0$, indicating a singularity in this region. We can see the same behavior for the torsion scalar (\ref{te1}), not revealing anything new, unlike some cases of charged black holes  in \cite{capozziello1}, where a new singularity emerged from the  analysis of  curvature scalar. At $ r = 0$, so the function $f(T)$ as the Lagrangian density $\mathcal{L}_{NED}$ diverge.

\subsubsection{b(r)=-4a(r)}
Taking  $N=4$, we have the particular case  $b(r)=-4a(r)$, where, after integration of the equation  (\ref{fT}) leads to 
\begin{eqnarray}
f(r)&&=16\int\exp\left[-\int\left(\frac{3(r-2M(r))(rM'(r)-M(r))}{4rM(r)(r-M(r))}\right)dr\right]\frac{1}{r^6}\Bigg\{r^3 M'^2(r) + 
 r M^2(r) \nonumber\\
 &&\left[-4 + 9 M'(r) + 6 M'^2(r) - 3rM''(r)\right]+r^2 M(r)\left[-M'(r)-6M'^2(r)+ rM''(r)\right]\nonumber\\
 &&+M^3(r)\left[5-10M'(r)+2rM''(r)\right]\Bigg\}dr\label{f2}\,.
\end{eqnarray}  
Using (\ref{LF}) into (\ref{F10}), one gets
\begin{eqnarray}
F^{10}(r)&=&\frac{[r-2M(r)]}{8qr^5M(r)[r-M(r)]}\exp\left[-\int\left(\frac{3(r-2M(r))(rM'(r)-M(r))}{4rM(r)(r-M(r))}\right)dr\right]\sqrt{1-\frac{2M(r)}{r}}\nonumber\\
&&\Bigg\{-96 M^6(r)+3r^6M'^2(r)+4r^2M^4(r)\left[-95+16M'(r)+24M'^2(r)-20rM''(r)\right]\nonumber\\
&&+32rM^5(r)\left[10-2M'(r)+rM''(r)\right]+2r^5M(r)\left[5M'(r)-18M'^2(r)+2rM''(r)\right]\nonumber\\
&&+12r^3M^3(r)\left[17+2M'(r)-16M'^2(r)+6rM''(r)\right]\nonumber\\
&&-r^4M^2(r)\left[45+40M'(r)-132M'^2(r)+28rM''(r)\right]\Bigg\}\label{F10-2}
\end{eqnarray}
By also taking the imposition (\ref{imposition1}) and using (\ref{L}) for $N=4$, one obtains the following equation 
\begin{eqnarray}
&&\left\{-3rM^2(r)+2M^3(r)-2r^2M(r)\left[-1 +M'(r)\right]M'(r)+r^3M'^2(r)\right\}\times\nonumber\\
&&\times\left[r^2-3rM(r)+2M^2(r)\right]^2=0\,.\label{si2}
\end{eqnarray}
The linear solutions in  $r$ implies an indetermination in  (\ref{imposition1}), and then, we discard these solutions. The unique acceptable solutions of (\ref{si2}) is
\begin{eqnarray}
M(r)=\frac{r}{2}-\frac{1}{2r}\sqrt{r^4-4e^{r_0}}\label{M2}\,.
\end{eqnarray} 
Remembering  (\ref{b=-Na}) and  (\ref{F10-2}), one then has the following solution 
\begin{eqnarray}
&&dS^2=\frac{1}{r^2}\sqrt{r^4-4e^{r_0}}dt^2-\frac{r^8}{\left(r^4-4e^{r_0}\right)^2}dr^2-r^2\left(d\theta ^2+\sin^2\theta d\phi^2\right)\label{s2}\,,\\
&&F^{10}(r)=-\frac{16e^{2r_0}}{qr^8}\left(r^4-4e^{r_0}\right)^{3/4}\label{F10-3}\,.
\end{eqnarray}

We can reconstruct the Lagrangian density  $\mathcal{L}_{NED}$ and the algebraic function  $f(T)$, as done before. From (\ref{s2}) and  (\ref{te}), one obtains  
\begin{eqnarray}
T(r)=-\frac{32e^{2r_0}}{r^{10}}\label{te2}\,.
\end{eqnarray}
Inverting this expression for obtaining $r(T)$ and using  (\ref{s2}) in (\ref{f2}), one gets 
\begin{eqnarray}
f(T)=-\frac{20}{7}\sqrt{2}e^{3r_0/5}\left(-T\right)^{7/10}\,.\label{f3}
\end{eqnarray}
One the same way, using (\ref{s2}) and (\ref{F10-3}) one has
\begin{eqnarray}
F(r)=-\frac{128e^{4r_0}}{q^2r^{10}}\label{F2}\,.
\end{eqnarray}
By inverting this expression for obtaining $r(F)$ and using $f(r)$ and (\ref{LF}) into (\ref{L}), one gets 
\begin{eqnarray}
\mathcal{L}_{NED}=\frac{5q^{7/5}e^{-4r_0/5}}{7\times 2^{9/10}}\left(-F\right)^{7/10}\label{L2}\,.
\end{eqnarray}

By using the components (\ref{tt}) and  (\ref{s2}), the curvature scalar is given by 
\begin{eqnarray}
R(r)=\frac{24e^{r_0}}{r^{10}}(r^4-4e^{r_0})\label{R2}\,.
\end{eqnarray}

We then see that the metric (\ref{s2}) is asymptotically Minkowskian. We has an event horizon in the region  $r=r_H=\left(4e^{r_0}\right)^{1/4}$. The component $F^{10}(r)$ into (\ref{F10-3}) decays with  $F^{10}\sim r^{-5}$ in the spacial infinite and vanishes at the horizon. 
\subsubsection{b(r)=-6a(r)}
Setting $N=6$, we have the particular case  $b(r)=-6a(r)$, and integrating the equation (\ref{fT}) one gets
\begin{eqnarray}
f(r)&&=8\int\exp\left[\int\left(\frac{5(M(r)-rM'(r))}{r(r-2M(r))\left(-1+\frac{1}{\left(1-\frac{2M(r)}{r}\right)^3}\right)}\right)dr\right]\frac{1}{r^8}\Bigg\{3r^5M'^2(r)\nonumber\\
&&-8M^5(r)\left[-7+14M'(r)-2rM''(r)\right]+3r^4M(r)\left[M'(r)-12M'^2(r)+rM''(r)\right]\nonumber\\
&&-40rM^4(r)\left[3 - 5M'(r)-2M'^2(r)+rM''(r)\right]-6r^3M^2(r)\left[4-3M'(r)-20M'^2(r)+3rM''(r)\right]\nonumber\\
&&+10r^2M^3(r)\left[9-12M'(r)-16M'^2(r)+4rM''(r)\right]\Bigg\}dr\label{f4}\,.
\end{eqnarray}  
Using (\ref{LF}) into (\ref{F10}) one gets
\begin{eqnarray}
F^{10}(r)&=&\frac{[r-2M(r)]^2}{4qr^8M(r)[3r^2-6rM(r)+4M^2(r)]}\int\exp\left[\int\left(\frac{5(M(r)-rM'(r))}{r(r-2M(r))\left(-1+\frac{1}{\left(1-\frac{2M(r)}{r}\right)^3}\right)}\right)dr\right]\nonumber\\
&&\sqrt{1-\frac{2M(r)}{r}}\Bigg\{1280 M^9(r)+5r^9M'^2(r)-128rM^8(r)\left[47-4M'(r)+2rM''(r)\right]\nonumber\\
&&+2r^8M(r)\left[19M'(r)-50M'^2(r)+3rM''(r)\right]-80r^5M^4(r)[50+38M'(r)-69M'^2(r)\nonumber\\
&&+13rM''(r)]+64r^2M^7(r)\left[191-12M'(r)-20M'^2(r)+16rM''(r)\right]\nonumber\\
&&+  80r^4M^5(r)\left[119+38M'(r)-84M'^2(r)+22rM''(r)\right]-r^7M^2(r)[115+376M'(r)\nonumber\\
&&-720M'^2(r)+72rM''(r)]+4r^6M^3(r)\left[245+376M'(r)-660M'^2(r)+92rM''(r)\right]\nonumber\\
&&-16r^3M^6(r)\left[867+56M'(r)-280M'^2(r)+112rM''(r)\right]\Bigg\}\label{F10-3}
\end{eqnarray}
Making use of the constraint (\ref{imposition1})  and using (\ref{L}) for $N=6$, one gets the following equation 
\begin{eqnarray}
&&\left[3r^2-6rM(r)+4M^2(r)\right]^2\Bigg\{8rM^3(r)-4M^4(r)-4r^3M(r)\left[-1+M'(r)\right]M'(r)\nonumber\\
&&+r^4M'^2(r)+r^2M^2(r)\left[-5-4M'(r)+4M'^2(r)\right]\Bigg\}=0\,.\label{si3}
\end{eqnarray}
The solution linear in  $r$ leads to a indetermination in  (\ref{imposition1}), and then we discard this solution and the complexes ones. The unique acceptable solution of (\ref{si3}) is  
\begin{eqnarray}
M(r)=\frac{r}{2}-\frac{1}{2r}\left(r^6-2r_0\right)^{1/3}\label{M3}\,.
\end{eqnarray} 
Remembering that  (\ref{b=-Na}) and  (\ref{F10-3}), one gets the following solution
\begin{eqnarray}
&&dS^2=\frac{1}{r^2}\left(r^6-2r_0\right)^{1/3}dt^2-\frac{r^{12}}{\left(r^6-2r_0\right)^2}dr^2-r^2\left(d\theta ^2+\sin^2\theta d\phi^2\right)\label{s3}\,,\\
&&F^{10}(r)=-\frac{4r_0^2}{qr^{12}}\left(r^6-2r_0\right)^{5/6}\label{F10-4}\,.
\end{eqnarray}

We can reconstruct the Lagrangian density   $\mathcal{L}_{NED}$ and the algebraic function  $f(T)$. From  (\ref{s3}) and  (\ref{te}), one gets  
\begin{eqnarray}
T(r)=-\frac{8r_0^2}{r^{14}}\label{te3}\,.
\end{eqnarray}
By inverting this expression for obtaining  $r(T)$ and using (\ref{s3}) in (\ref{f4}) we obtain 
\begin{eqnarray}
f(T)=-\frac{28}{9}2^{1/14}r_0^{5/7}\left(-T\right)^{9/14}\,.\label{f5}
\end{eqnarray}
On the same way, using (\ref{s3}) and  (\ref{F10-4}) one has 
\begin{eqnarray}
F(r)=-\frac{8r_0^4}{q^2r^{14}}\label{F3}\,.
\end{eqnarray}
Also, inverting this expression for obtaining  $r(F)$ and using  $f(r)$ and (\ref{LF}) into (\ref{L}) we get
\begin{eqnarray}
\mathcal{L}_{NED}=\frac{7\times 2^{1/14}q^{9/7}}{9r_0^{4/7}}\left(-F\right)^{9/14}\label{L3}\,.
\end{eqnarray}

Making use of the components (\ref{tt}) and (\ref{s3}), the curvature scalar is given by 
\begin{eqnarray}
R(r)=\frac{20r_0}{r^{14}}{(r^6-2r_0)}\label{R3}\,.
\end{eqnarray}

This solution is asymptotically Minkowskian and has event horizon at $r=r_H=\left(¨2r_0\right)^{1/6}$. The component  $F^{10}(r)$ of the  Maxwell tensor behaves as  $F^{10}\sim r^{-7}$ in the infinite space, also vanishing at the event horizon.
\par 
{\bf Important observations}. We can now note a certain logic for the solutions  $b=-2a,-4a,-6a$. Let us generalize the reconstructions of (\ref{f1}), (\ref{f3}) and  (\ref{f5}), then, as Lagrangian densities (\ref{L1}), (\ref{L2}) and  (\ref{L3}) for
\begin{eqnarray}
f(T)=f_0 \left(-T\right)^{(N+3)/[2(N+1)]}\;,\;\mathcal{L}_{NED}=\mathcal{L}_0\left(-F\right)^{(N+3)/[2(N+1)]}\label{fN}\,,
\end{eqnarray}
with an even $N$. Only a particular case is admitted here, when $N=1$ we recover the TT with the Maxwell Lagrangian density. All of these solutions are asymptotically Minkowskian. The effect of the electric field vanishes at spatial infinity. The integration constant $r_0$ appearing in all solutions should be that depend implicitly on the mass and charge of the black hole, so that if we take the limit  $\lim _{m,q\rightarrow 0}r_0\rightarrow 0$, thus returning to the Minkowskian vacuum.
\par 
When we choose an odd $N$ in (\ref{b=-Na}), on very complicated differential equations where the analytical solutions are almost impossible to be obtained. Thus, we do not take them into account
\subsection{Solutions with $b(r)=Na(r)$}\label{subsec5.2}
By the symmetry, we can take the same equations (\ref{fT})-(\ref{LF}), but setting  $N\rightarrow -N$ in all of them. Let us see some particular cases for fix values of $N$.

\subsubsection{b(r)=a(r)}
Taking the particular case $N=1$, i.e, $b(r)=a(r)$ and performing the same process as in the above subsections, we obtain the following solution 
\begin{eqnarray}
&&dS^2=\left(e^{r_0/2}r-1\right)^{-2}\left[dt^2-dr^2\right]-r^2\left(d\theta ^2+\sin^2\theta d\phi^2\right)\label{s4}\,,\\
&&F^{10}(r)=-\frac{r\left(e^{r_0/2}-2\right)^2}{q\left(e^{r_0/2}r-2\right)^3}\left(2-4e^{r_0/2}r+e^{r_0}r^2\right)\left(e^{r_0/2}r-1\right)^2e^{r_0/2}\label{F10-5}\,.
\end{eqnarray}
\par 
This one is the charged and asymptotic non-flat black hole solution, with an degenerated event horizon in  $r=r_H=e^{-r_0/2}$. The component $F^{10}(r)$ diverging at spacial space, what may happen for asymptotically non-flat solutions, due to the possibility of non-null field at spatial infinity continue twisting the space-time.
\subsubsection{b(r)=2a(r)}
Taking into account the particular case $N=2$, i.e, $b(r)=2a(r)$  and making the same process as in the previous subsections, we find the following solution 
\begin{eqnarray}
&&dS^2=\left(1-\frac{2r^2}{r_0}\right)^{-1}dt^2-\left(1-\frac{2r^2}{r_0}\right)^{-2}dr^2-r^2\left(d\theta ^2+\sin^2\theta d\phi^2\right)\label{s5}\,,\\
&&F^{10}(r)=-\frac{4r}{qr_0^2}\left(1-\frac{2r^2}{r_0}\right)^{-3/2}\label{F10-6}\,.
\end{eqnarray}
This solution not seems physically acceptable, due to the fact that for $r^2>r_0/2$ one has the following signature for the metric $(----)$. This should be the exterior region to the horizon, but for which the Killing vectors do not change. Here, the Killing vector are only space-like for the exterior region to the event horizon, and this does not permit to distinguish between the physical space and time. At least, it seems for us that this solution has to be discarded.
\subsubsection{b(r)=4a(r)}
Taking the particular case $N=4$, i.e, $b(r)=4a(r)$ and making the same process as performed in the previous subsections, one gets the following solution 
\begin{eqnarray}
&&dS^2=\left(1+4e^{r_0}r^4\right)^{-1/2}dt^2-\left(1+4e^{r_0}r^4\right)^{-2}dr^2-r^2\left(d\theta ^2+\sin^2\theta d\phi^2\right)\label{s6}\,,\\
&&F^{10}(r)=-\frac{16e^{2r_0}r^3}{q}\left(1+4e^{r_0}r^4\right)^{5/4}\label{F10-7}\,.
\end{eqnarray}
This solution does not possess event horizon, being a solution of a geometry with atypical non vanish torsion.

\subsubsection{b(r)=6a(r)}

Taking the particular case $N=6$, i.e,  $b(r)=6a(r)$ and making use of the same process as in the above subsection, we find the following solution
\begin{eqnarray}
&&dS^2=e^{r_0}\left(e^{3r_0}+2r^6\right)^{-1/3}dt^2-e^{6r_0}\left(e^{3r_0}+2r^6\right)^{-2}dr^2-r^2\left(d\theta ^2+\sin^2\theta d\phi^2\right)\label{s6}\,,\\
&&F^{10}(r)=-\frac{4e^{-9r_0}r^5}{q}\left(e^{3r_0}+2r^6\right)^{7/6}\label{F10-7}\,.
\end{eqnarray}
In the same way as in the  previous solution, this one also does not possess event horizon, being only a non-null torsion, but without any interest.  
\par 
{\bf Important observations}. We can conclude that the cases with  $N=3,5,7,...$ lead to more complicated differential equations that do not yield atypical solutions. However with $N=4,6,...$ there is no event horizon, being a non interesting geometries.

\subsection{Born-Infeld-type solutions with $b(r)\neq -a(r)$}\label{subsec5.3}

Let us start taking the Lagrangian density of  BI (\ref{lBI}) and the function $f_T(r)$ into (\ref{fT}), which is independent of the matter content. Now we solve the equation (\ref{eq1}), obtaining  $f(r)$
\begin{eqnarray}
f(r)&&=2\Bigg\{\frac{4q^2\beta_{BI}}{r^2\sqrt{q^2+4\beta^2_{BI}r^4}}-8\beta^2_{BI}\left[1-\frac{2r^2\beta_{BI}}{r^2\sqrt{q^2+4\beta^2_{BI}r^4}}\right]+\frac{f_T(r)e^{-b}}{r}\left(a'+b'\right)\nonumber\\
&&-\frac{f_T(r)e^{-b}}{r^2}\left[\left(2+ra'\right)\left(e^{b/2}-1\right)+rb'\right]\Bigg\}\label{f6}\,.
\end{eqnarray}
Now, the new condition, for the solution being consistent, is
\begin{eqnarray}
f_T(r)=\frac{df(r)}{dr}\left(\frac{dT(r)}{dr}\right)^{-1}\label{imposition2}\,.
\end{eqnarray}
For this condition being satisfied, and using (\ref{eq3}), we must have  
\begin{eqnarray}
8q^2\beta_{BI}\left(2-2e^{b/2}+ra'\right)\left(a'+b'\right)=0\label{eqBI}\,.
\end{eqnarray}
As $b(r)=-a(r)+b_0$ leads to $f_T(r)=f_1\in\Re$, we will not use this solution here. By solving the equation  (\ref{eqBI}), we obtain the following possibility 
\begin{eqnarray}
e^{b(r)}=\left[\frac{1}{2}\left(2+ra'\right)\right]^2\label{bBI}\,.
\end{eqnarray}
Now the function $a(r)$ has to be obtained by integrating the equation (\ref{eq3}).  We will do this as follows. First we isolate  $f_T(r)$. But as  $f_T(r)$ is already given in  (\ref{fT}), we extract the logarithm of the expression and derive, for then equal from the argument of the integration of the exponential term in (\ref{fT}). This leads to the following differential equation 
\begin{eqnarray}
\frac{1}{(2+ra')}\left[2\frac{a''}{a'}+\frac{16r^3\beta_{BI}^2-(q^2-4r^4\beta^2_{BI})a'}{q^2+4\beta^2_{BI}r^4}\right]=0\label{eq3-BI}\,.
\end{eqnarray} 
The general solution of this equation is given by 
\begin{eqnarray}
e^{a(r)}=\exp\left[2\int\frac{dr}{2r_0\sqrt{q^2+4\beta^2_{BI}r^4}-r}\right]\label{aBI}\,.
\end{eqnarray}
With (\ref{aBI}) and  (\ref{bBI}), one gets 
\begin{eqnarray}
f_T(r)=\frac{1}{r^2}\sqrt{q^2+4\beta^2_{BI}r^4}\label{fTBI}\,.
\end{eqnarray}
here we solve the equation of motion (\ref{eq1}) for determining $f(r)$, which, substituting into (\ref{eq3}) lead to the following integration constant $r_0=1/[2q\sqrt{2\beta_{BI}}]$. At this point, all the  equation and conditions of consistency are satisfied. We then have the following solution 
\begin{eqnarray}
&&dS^2=\exp\left[2\int\frac{dr}{\frac{\sqrt{q^2+4\beta^2_{BI}r^4}}{q\sqrt{2\beta_{BI}}}-r}\right]dt^2-\frac{2(q^2+4\beta^2_{BI}r^4)}{\left(\sqrt{2q^2+8\beta^2_{BI}r^4}-2qr\sqrt{\beta_{BI}}\right)^2}dr^2\nonumber\\
&&-r^2\left(d\theta^2 +\sin^2\theta d\phi^2\right)\label{s7}\,,\\
&&F^{10}(r)=\frac{2q\beta_{BI}}{q^2+4\beta_{BI}^2r^4}\left(qr\sqrt{2\beta_{BI}}-\sqrt{q^2+4\beta^2_{BI}r^4}\right)\exp\left[-\int\frac{dr}{\frac{\sqrt{q^2+4\beta^2_{BI}r^4}}{q\sqrt{2\beta_{BI}}}-r}\right]\,.\label{F10-8}
\end{eqnarray}
The interesting point here is that the Lagrangian density is the  BI in (\ref{lBI}). The solutions with $b(r)\neq -a(r)$ also have been obtained for $f(R)$ theory \cite{hollenstein}, but with a strong restriction, where  $a+b=\ln[k_0r^{l}],l,k_0\in\Re$. 
\par 
We can reconstruct the function   $f(T)$ here. The torsion scalar (\ref{te}), with (\ref{s7}), is given by  
\begin{eqnarray}
T(r)=-\frac{4q^2\beta_{BI}}{q^2+4\beta^2_{BI}r^4}\label{teBI}\,.
\end{eqnarray}
The function $f(r)$ obtained from (\ref{eq1}) is given by 
\begin{eqnarray}
f(r)=-16\beta^2_{BI}\left[1-\frac{2\beta_{BI}r^2}{\sqrt{q^2+4\beta^2_{BI}r^4}}\right]\label{f7}\,.
\end{eqnarray} 
By inverting the expression (\ref{teBI}) for obtaining $r(T)$  and substituting it into (\ref{f7}), one gets the following function 
\begin{eqnarray}
f(T)=-16\beta^2_{BI}\left[1-\sqrt{1+\frac{T}{4\beta_{BI}}}\right]\label{f8}\,.
\end{eqnarray}
We have a solution of a charged black hole, spherically symmetric and static,  and asymptotically Minkowskian, because of  $e^{a(r)},e^{b(r)}\rightarrow 1$ for $r\rightarrow +\infty$.  The expression of the curvature scalar is too large to be written here, but in the limit of infinite space, it vanishes, and for the limit of the radial coordinate origin, it diverges, indicating a singularity. On the other hand   $F^{10}(r)$ goes to zero at spatial infinity and is regular, with the value $-2\beta_{BI}$, at the origin $r=0$. The solution possesses two event horizons in

\begin{eqnarray}
r_{\pm}=\frac{q}{2\sqrt{\beta_{BI}}}\sqrt{1\pm \frac{1}{q}\sqrt{q^2-4}}\,\label{hBI}.
\end{eqnarray}
The extreme limit is given for $q=2$, where  $r_+=r_-=1/\sqrt{\beta_{BI}}$.

\section{Conclusion}\label{sec6}
Here we reconsidered the original idea of BI for a nonlinear Lagrangian density in $F$ implying that for certain relationship  between the metric functions $ a(r) $ and $ b(r) $, a function $ f(T) $ with the same functional dependence of the NED action. We can re-obtain so the famous BI solutions with cosmological constant and the one of Reissner-Nordstrom-AdS,  of GR.
\par 
A very strong result is established by no-go theorem of the section \ref{sec4} where we only can have solutions in the $f(T)$ theory with non-linear terms in $ T $, when we break down the famous symmetry $b(r)=-a(r)$ and $ \mathcal{T}_{0}^{\;\;0} =\mathcal{T}_{1}^{\;\;1}$ for the tetrads (\ref{ndtetrad}).
\par  
We  performed a powerful algorithm for solving differential equations of this theory. We see that still can be addressed more new cases in which the relationship between the metric functions are different from those used here. It is also possible to consider metric functions such that the solution is a regular black hole in $f(T)$ Gravity. This is currently being done by us in another work \cite{rodrigues4}. The new solutions that merit to be  pointed out are charged and asymptotically flat  black holes solutions whose Lagrangian densities and action functions are given by $\mathcal{L}_{NED}=\mathcal{L}_0\left(-F\right)^{(N+3)/[2(N+1)]}$ and $f(T)=f_0 \left(-T\right)^{(N+3)/[2(N+1)]}$, when  $b(r)=-Na(r)$ with an even $N$.
\par 
Another new solution obtained here is when we consider the no-go theorem and the Lagrangian density BI. For this model the solution is asymptotically flat with two or one horizon (extreme case). Interestingly, the functional form of $ f(T) $ is identical to the Lagrangian density of BI, but with $ F\rightarrow T $, which recovers the original idea of Ferraro and Fiorini in \cite{ferraro1}, but concerning  black hole here.

\vspace{1cm}

{\bf Acknowledgement}: Ednaldo L. B. Junior thanks CAPES for financial support. Manuel E. Rodrigues  
thanks UFPA, Edital 04/2014 PROPESP, and CNPq, Edital MCTI/CNPQ/Universal 14/2014,  for partial financial support.
Mahouton J. S. Houndjo thanks ENS-Natitingou for partial financial support.

%

\end{document}